\documentclass[twocolumn, amsmath,amssymb,prl]{revtex4}
\usepackage{latexsym}
\usepackage{graphicx}
\usepackage{psfig}
\usepackage{color}
\usepackage{gensymb}
\usepackage{verbatim}
\usepackage{multirow}
\usepackage{physics}
\usepackage{hyperref}
\usepackage[letterpaper, margin=1in]{geometry}
\makeatletter
\DeclareRobustCommand\citenum
   {\begingroup
     \NAT@swatrue\let\NAT@ctype\z@\NAT@parfalse\let\textsuperscript\relax
     \NAT@citexnum[][]}
\makeatother

\usepackage{dcolumn}
\usepackage{bm}
\usepackage[utf8]{inputenc}
\usepackage[T1]{fontenc}
\usepackage{mathptmx}

\begin{document}

\preprint{AIP/123-QED}
\title[Epitaxial Superconductor-Semiconductor Two-Dimensional Systems for Superconducting Quantum Circuits]{Epitaxial Superconductor-Semiconductor Two-Dimensional Systems for Superconducting Quantum Circuits}

\author{Joseph~O'Connell~Yuan}
\author{Kaushini~S.~Wickramasinghe}
\author{William~M.~Strickland}
\author{Matthieu~C.~Dartiailh}
\author{Kasra~Sardashti}
\author{Mehdi~Hatefipour}
\author{Javad~Shabani}

\affiliation{Center for Quantum Phenomena, Department of Physics, New York University, NY 10003, USA}

\date{\today}

\begin{abstract}
Qubits on solid state devices could potentially provide the rapid control necessary for developing scalable quantum information processors.  Materials innovation and design breakthroughs have increased functionality and coherence of qubits substantially over the past two decades. Here we show by improving interface between InAs as a semiconductor and Al as a superconductor, one can reliably fabricate voltage-controlled Josephson junction field effect transistor (JJ-FET) that can be used as tunable qubits, resonators, and coupler switches. We find that band gap engineering is crucial in realizing a two-dimensional electron gas near the surface. In addition, we show how the coupling between the semiconductor layer and the superconducting contacts can affect qubit properties. We present the anharmonicity and coupling strengths from one and two-photon absorption in a quantum two level system fabricated with a JJ-FET.

This article may be downloaded for personal use only. Any other use requires prior permission of the author and AIP Publishing. This article appeared in Journal of Vacuum Science \& Technology A 39, 033407 (2021) and may be found at \hyperref[https://doi.org/10.1116/6.0000918]{https://doi.org/10.1116/6.0000918}.


\end{abstract}

\maketitle

\section{Background}
Recent advancements in materials synthesis and engineering have enabled the design and fabrication of a variety of novel quantum devices including sensors~\cite{pirandola_advances_2018, sensors1}, low noise amplifiers~\cite{macklin_nearquantum-limited_2015, amplifiers1, amplifiers2, amplifiers3}, and highly coherent qubits for quantum computation~\cite{arute_quantum_2019, bravyi_quantum_2020}. In the context of quantum information processors, qubit coherence is crucial in bringing error rates below a threshold at which error correcting algorithms may allow for the realization of a fault tolerant quantum computer~\cite{Fowler2009, DiVincenzo_2009, Fowler2012}. As it has been demonstrated in superconducting qubits, design and materials advances have improved the lifetimes from nanoseconds to hundreds of microseconds~\cite{devoret_implementing_2004, oliver_welander_2013, currentstateofplay, Krantz2019}. Recent experiments~\cite{barends_superconducting_2014, kelly_state_2015, Sheldon2016, fidelity2019, fidelity2020, jurcevic2020demonstration} have also demonstrated single- and two-qubit gate operation fidelities exceeding 99\%.



In scaling the size and power of current quantum information processors, it is necessary to fabricate a large number of highly coherent qubits on a single chip. However, there will be always a trade-off in coupling qubits for gate operations and isolating qubits well enough from each other when not in use as to preserve their encoded information. This eventually becomes an issue in fixed frequency qubits where closely spaced energy levels prevents fast, high-fidelity control. One approach to mitigate this issue resorts to slower qubit driving at the expense of greater exposure to noise. As operations must be performed on the order of the qubit lifetimes, however, this could not be viable. 

A second approach is to work with qubits which can rapidly be tuned in and out of resonance with each other, keeping the system in a relatively noise-resistant configuration without sacrificing operation times. One approach is to use semiconductor based circuit elements in which the density and conductivity in the semiconductor region can be tuned using an applied gate voltage~\cite{deLange2015,Larsen_PRL}. This is manifested in the Josephson junction field effect transistor (JJ-FET), a Josephson junction (JJ) in which the insulating weak-link is replaced by a semiconductor region. JJ-FETs rely on the proximity effect, where superconducting Cooper pairs tunnel through the semiconductor region, allowing the supercurrent through the junction to be tuned via a gate voltage applied to the semiconductor region. JJ-FETs have been realized with success in various materials systems, including III-V quantum wells~\cite{Billy2019, Casparis2018, Feng}, graphene~\cite{calado_ballistic_2015, Park_2018, rodanlegrain2020highly}, and nanowires~\cite{Larsen_PRL, Xiang_2006}. Gate voltage tuning of the JJ-FET should effectively carry no current and only sustain a voltage level, meaning there is little to no power loss through this circuit element. Pure gate voltage tunability also means that JJ-FETs can work in the presence of a magnetic field as opposed to superconducting quantum interference devices (SQUIDs). In this work, we present the development of the JJ-FET, starting with epitaxial growth of Al on InAs, and then expanding on their fabrication and use in quantum systems for single photon measurements.

\section{Epitaxial I\lowercase{n}A\lowercase{s}-A\lowercase{l} structures}

\begin{figure*}[htp]
    \centering
    \includegraphics[width=\textwidth]{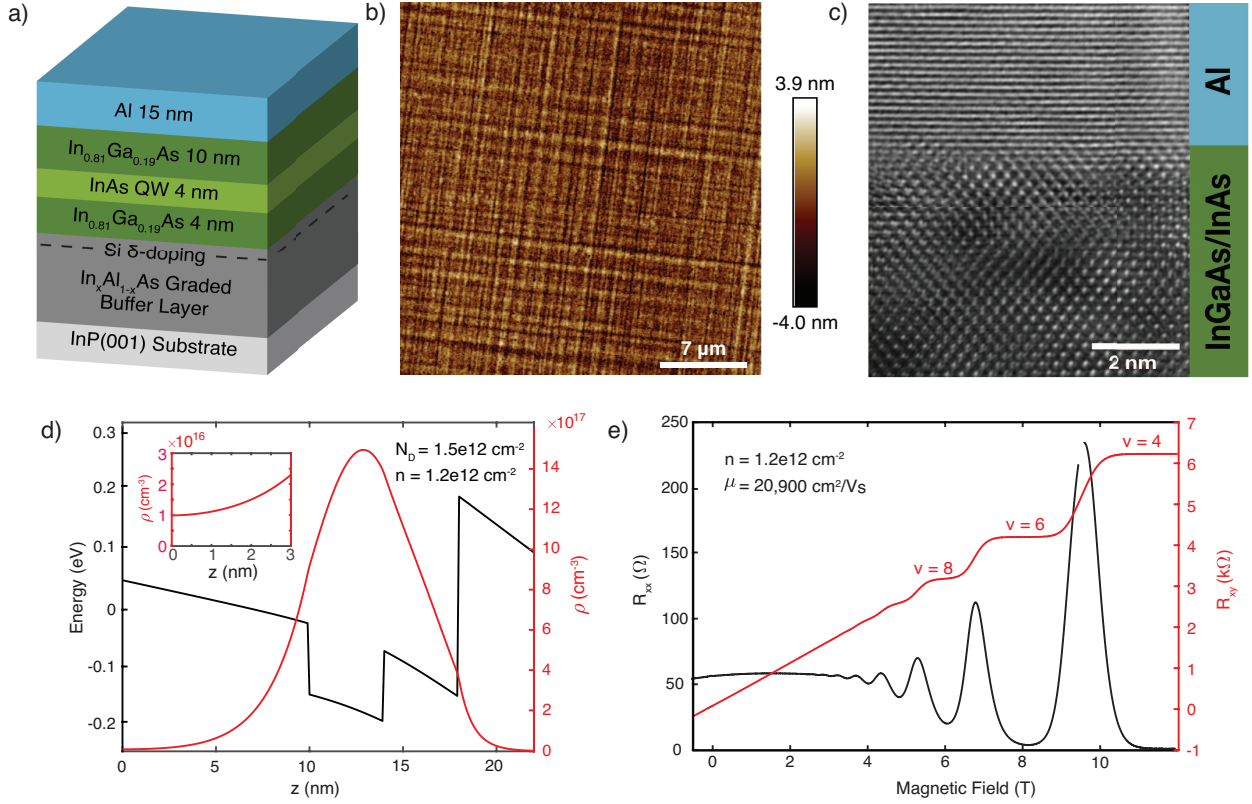}
    \caption{(a) Schematic of the Al/InAs structure. (b) Atomic force microscopy image of the aluminum surface shows an average surface roughness of 0.8 nm. (c) High resolution transmission electron microscopy image showing the sharp interface between the Al thin film and the InGaAs top layer. (d) Self consistent Schr$\"{o}$dinger-Poisson calculations for the electron density $\rho$ (red) and conduction band energy (black) of the III/V heterostructure for a Si $\delta$-doping $N_D = 1.5\times 10^{12}$ as a function of the thickness $z$ measured from the In$_{0.81}$Ga$_{0.19}$As top barrier surface down towards the substrate. The 2D electron density is $n = 1.2 \times 10^{12}$. The inset (top left) shows the electron density near the Al/InGaAs interface. (e) Magnetotransport measurement of an InAs quantum well with a 10 nm top layer of InGaAs with an electron density of $1.2\times 10^{12}$ cm$^{-2}$ and a mobility of $\mu = 20,900$ cm$^{2}$/Vs. Quantum Hall states at filling factors $\nu =$4, 6, 8 are visible.}
   \label{fig:FFG0}
\end{figure*}

We use molecular beam epitaxy, an ultrahigh vacuum deposition technique for ordered growth of crystalline thin films~\cite{Mccray2007MBEDA}. Recently it was shown that it is possible to grow epitaxial layers of Al on an InGaAs ternary layer under optimal growth conditions and the appropriate selection of metallic phases in order to suppress the strong tendency for island formation and film agglomeration during growth~\cite{Shabani2016, sarney_2020_metallization, sarney2018_reactivity}. A schematic of the materials structure we study is shown in Fig.~\ref{fig:FFG0}(a). Since the lattice parameter of the InAs active region is greater than that of the InP substrate (with a lattice mismatch of about 3$\%$), the formation of misfit and threading dislocations is unavoidable. To alleviate this, we attempt pseudomorphic growth in which we use a step graded buffer where the composition of the In$_{x}$Al$_{1-x}$As buffer layer ranges from $x=0.52$ to $x=0.81$ in steps of $\Delta x = 0.02$ every 50~nm. This results in a cross-hatched surface morphology where dislocations meet, which is confirmed by an atomic force micrograph of the sample shown in Fig.~\ref{fig:FFG0}(b). In order to deposit Al layer-by-layer, the substrate is cooled to sub-zero temperatures to promote the growth of Al (111)~\cite{Shabani2016}. Fig.~\ref{fig:FFG0}(c) shows a cross-sectional high resolution transmission electron microscopy (HR-TEM) image which reveals an atomically flat, epitaxial contact made by the Al layer on the In$_{0.81}$Ga$_{0.19}$As top barrier.

A complication involving a layer of InAs at the surface is the negative Schottky barrier and its Fermi level pinning. The resulting charge accumulation at the surface makes design and control of the electron density difficult. By adding gallium to an In$_x$Ga$_{1-x}$As ternary top barrier layer, the Fermi level pinning can be tuned to where it crosses the conduction band near $x = 0.85$. We use In$_{0.81}$Ga$_{0.19}$As which has a small, positive Schottky barrier~\cite{adachi2009_schottky} of 0.04 eV. This Schottky barrier is small enough for carriers to be introduced in the quantum well through backside modulation doping 6 nm below the quantum well. In addition, the electron effective mass of In$_{0.81}$Ga$_{0.19}$As is $m_{e}$ = 0.03$m_{0}$, where $m_{0}$ is the bare mass, and is comparable to the electron effective mass of InAs which is $m_{e}$ = 0.023$m_{0}$\cite{Yuan2020}. This results in a quantum well structure whose conduction band edge supports the confinement of the wavefunction both in the In$_{0.81}$Ga$_{0.19}$As and InAs layers, but also allows the wavefunction to decay slowly toward the surface, allowing for a sufficiently strong overlap with the superconductor.

We use 1D Poisson~\cite{Tan1990, Snider1990}, a calculator which self-consistently solves the Schr$\"{o}$dinger and Poisson equations in 1D. The charge density and conduction band edge of the structure described above are shown in Fig.~\ref{fig:FFG0}(d). The band gaps for InAs, In$_{0.81}$Ga$_{0.19}$As, and In$_{0.81}$Al$_{0.19}$As used in the calculator 0.372 eV, 0.520 eV, and 0.880 eV respectively. The band offset of InAs relative to InP is -0.670 eV, the band offset of In$_{0.81}$Ga$_{0.19}$As relative to InP is -0.550 eV, and the band offset of In$_{0.81}$Al$_{0.19}$As relative to InP is -0.210 eV. The electron effective masses written as proportions of the bare electron mass used for InAs and In$_{0.81}$Ga$_{0.19}$As are mentioned in the previous paragraph, and that of In$_{0.81}$Al$_{0.19}$As is 0.070. The thickness $z$ is measured from the surface of the In$_{0.81}$Ga$_{0.19}$As top barrier down towards the substrate. The Fermi level of the Al layer lies at 0 eV and calculations are conducted for a temperature of 10 K. These calculations assume the Fermi level pinning is fixed by the semiconductor surface. The charge density and conduction band structure are simulated with a sheet charge $N_D = 1.5 \times 10^{12}$ cm$^{-2}$, assuming full ionization of donors. 

While the strong coupling between the Al layer and the InAs two-dimensional electron gas (2DEG) is of central importance for JJ-FET fabrication, the electron mobility is inversely affected when the InAs 2DEG is placed close to the surface immediately under the Al layer. The quantum well design should consider these two competing factors. It has been shown that a 10~nm top InGaAs layer yields high mobility while maintaining good coupling to the superconductor, as seen by a nonzero electron density present at the surface, shown in the inset of Fig.~\ref{fig:FFG0}(d) by the simulated electron density in the top 3 nm of the In$_{0.81}$Ga$_{0.19}$As top barrier. The transport properties of the 2DEG in this surface quantum well structure can be characterized in the presence of a perpendicular magnetic field. Figure~\ref{fig:FFG0}e shows longitudinal and Hall data where quantum Hall states are well developed above 5 T. All measurements are performed in a cryogen-free superconducting magnet system at a sample temperature of 1 K. The electron density of $n = 1.2 \times 10^{12}$ cm$^{-2}$ and electron mobility of 20,900 cm$^{2}$/Vs are achieved with a Si doping of $N_D = 1.5 \times 10^{12}$ cm$^{-2}$ delta doping. This is in close agreement with self-consistent calculations shown in Fig.~\ref{fig:FFG0}(d). The effects of various scattering mechanisms have been studied, and while surface roughness, background, ionized impurity scattering and alloy scattering all play a role, it is found that surface scattering dominates transport properties~\cite{Kaushini2018}.

\section{Josephson junction field effect transistors}

Josephson tunnel junctions are the basic building block of superconducting qubits as they provide the essential nonlinearity (in the form of a nonlinear inductor) that allows the formation of a distinct two level quantum system. Josephson junction properties can be probed by DC or microwave measurements. In this section, we discuss fabrication of the JJ-FET and their DC characteristics.

\begin{figure}
    \centering
    \includegraphics[width=.5\textwidth]{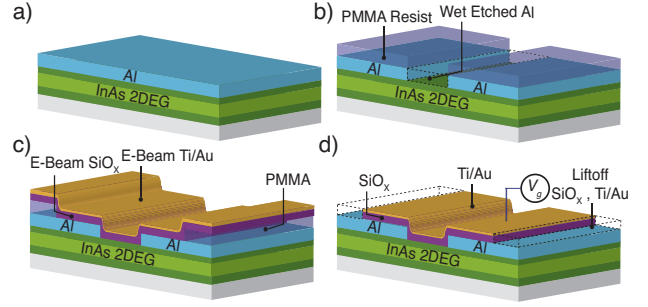}
   \caption{Fabrication Procedure (a) InAs channel near the surface which hosts the 2DEG, contacted epitaxially with Al. (b) Polymethyl methacrylate (PMMA) E-beam resist is spun over the surface, then patterned with E-beam lithography. A thin gap of Al is then selectively wet etched, leaving a thin gap between the Al contacts. The length of the junctions (in this work $L = 100$ nm) is defined by the separation of the two Al leads. (c) After another round of lithographic patterning, SiO$_x$ and Ti/Au are deposited via E-Beam deposition for the gate dielectric and electrode respectively. (d) Lift-off of the SiO$_x$ and Ti/Au, leaving the gate stack over the junction gap able to apply a gate voltage $V_g$.}
    \label{fig:FFG2}
\end{figure}

Starting with the epitaxial InAs-Al wafer, we can fabricate JJ-FETs using top-down electron beam (E-beam) lithographic patterning. The fabrication steps involve chemical wet etches of Transene Type D and a III-V etchant which target the Al and III-V layers respectively. The III-V etchant is a solution of phosphoric acid, hydrogen peroxide, and water solution created at the time of etching in a ratio of 1:1:80 respectively. We define a small gap where aluminum is etched away leaving two Al leads separated by a length $L =$ 100 nm, leaving the primary electrical path through the 2DEG below. After another round of lithographic patterning, a 50~nm SiO$_x$ gate dielectric and a 5~nm of titanium and 60~nm of gold are deposited using E-beam deposition. These two layers are then lifted off in 70~C acetone leaving only the metal deposited into the patterned areas and is used as gate electrodes. A schematic of the JJ-FET device is shown in Fig.~\ref{fig:FFG2}(a) -~\ref{fig:FFG2}(d) detailing the fabrication procedure.

\begin{figure}
    \centering
    \includegraphics[width=.45\textwidth]{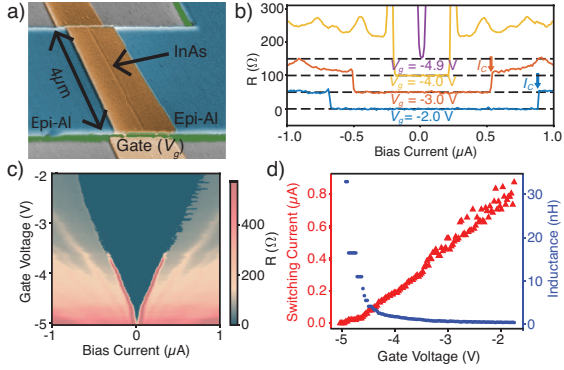}
   \caption{(a) False color scanning electron microscope image of a JJ-FET with a junction width of 100 nm. (b) Differential resistance for the 100~nm JJ-FET showing the critical current $I_C$ when the junction switches from zero-voltage to normal state. Shown are line cuts at different gate voltages $V_g$ which are vertically offset in 50 $\Omega$ steps. (c) 2D plot of differential resistance vs. applied bias current for a 100 nm JJ-FET. The dark blue represents zero resistance. (d) Switching current of the junction shown in red with triangle markers and tied to values on left y-axis.  Junction inductance for the corresponding switching current shown in blue with circle markers and tied to values on the right y-axis.  Both quantities shown as a function of gate voltage.  Inductance $L$ is found as a function of critical current from $L = \Phi_0/2\pi I_C$.}
    \label{fig:FFG3}
\end{figure}


There are three significant length scales concerning JJ-FETs: the mean free path in the semiconductor, $l_e$, the superconducting coherence length, $\xi_{0}$, and the junction length, $L$. The mean free path can be deduced from the 2DEG mobility and electron density. In the case of the sample shown in Fig.~\ref{fig:FFG0}(e), we find $l_e$ = 87 nm. We measure an Al superconducting critical temperature of about $T_C =$ 1.5 K, from which we can estimate the superconducting gap using $\Delta = 1.75 k_{B} T_{C}$ to be close to 231 $\mu$V. The superconducting coherence length is given by $\xi_{0} =\hbar v_f /\pi \Delta$  which yields $\xi_{0}= 635$ nm. 

For our junctions of length $L = 100$~nm we have $L \sim l_{e} < \xi_{0}$, which places our JJ-FETs in the dirty, short ballistic regime~\cite{Ambegaokar1963, Likharev1979}. In this regime, the product of the normal resistance $R_N$ and the supercurrent $I_C$ is theoretically expected to be $\pi \Delta$ for a fully transparent interface. Figure ~\ref{fig:FFG3}(b) shows the differential resistance versus DC bias current for several gate voltage values applied to our JJ-FET. The traces in Fig. ~\ref{fig:FFG3}(b) are offset from the bottom in 50 $\Omega$ steps to clearly show the zero resistance state.  The dashed lines indicate the relative 0 $\Omega$ level for each trace.  The critical current, $I_C$, can be identified by the transition of the zero resistance state to a state of finite resistance as marked at a couple example points where it leaves the superconducting state. The selected examples demonstrate the gate voltages showing the diminishing $I_C$ as the gate is driven negative.  We find the normal resistance by applying a magnetic field above the critical field or raising the temperature above $T_{C}$. The product of the normal resistance and the critical current in this JJ-FET reaches a maximum of $I_{C}R_{N} = 0.29 \Delta$. It is smaller than nominally identical devices with ALD deposited oxide~\cite{Billy2019}, being $I_{C}R_{N} = 2.17 \Delta$. 

For comparison Table~\ref{table:1} shows critical values of a Josephson junction, namely the junction length $L$, the critical current $I_{C}$, the normal resistance $R_N$, and the product $I_CR_N$ normalized to the superconducting gap $\Delta$ for a variety of nanowire and quantum well devices. While there is a clear advantage in epitaxy, we find in our samples that the SiOx liftoff process can damage the induced gap, represented by the product $I_C R_N$, as compared to conformal ALD oxide deposition~\cite{Billy2019}. Also included are results from different host materials~\cite{mayer_superconducting_2020, ke_ballistic_2019, wiedenmann_4-periodic_2016, Pallecchi2008}, including InAs$_{0.50}$Sb$_{0.50}$, InSb, HgTe, and carbon nanotubes (CNT). While for qubit applications, only $I_C$ enters the frequency of the qubit, $I_CR_N$ is an important parameter as it characterizes the strength of the induced gap in the semiconductor and Andreev bound states that govern the supercurrent in the junction \cite{Billy2019}.

As mentioned, the most noteworthy feature of our JJ-FET is the gate-voltage tunability of the supercurrent. Fig.~\ref{fig:FFG3}(c) shows the differential resistance in color as current bias and gate voltage are changed. The critical current is tuned from $\sim 1$ $\mu$A to 0 $\mu$A by sweeping the applied gate voltage, $V_{g}$, from -2~V to -5~V. This voltage scales with the thickness of dielectric layer. We again note that this tunability does not require any power consumption as there is virtually no leakage current flowing to the gate contact as the gate voltage is varied in this range.  The inductance of the junction is dependant on the critical current by $L = \Phi_0/2\pi I_C$. For the each gate voltage we extract the critical current and then calculate the resulting inductance as shown in Fig.~\ref{fig:FFG3}(d).

\begin{table*}[t]
\centering
\begin{tabular}{|| c | c | c | c | c | c | c | c | c ||} 
 \hline
 Identifier & Materials & Dimension & Technique & $L$ (nm) & $I_C$ ($\mu$A) & $R_N$ ($\Omega$) & $I_C R_N/\Delta$\\ 
 \hline\hline
 Our Work w/ Oxide Lift-off & InAs/Al & QW & Epitaxy & 100 & 0.9 & 74 & 0.29 \\
 \hline  
 Our Work with ALD Oxide [\citenum{Billy2019}] & InAs/Al & QW & Epitaxy & 100 & 5.00 & 102 & 2.2 \\
 \hline
 Delfanazari \textit{et al.} 2017 [\citenum{delfanazari_2017}] & InAs/Nb & QW & Sputtering  & 850 & 2.00 & 800 &  1.45 \\
 \hline
  Nitta \textit{et al.} 1992 [\citenum{Nitta92}] & InAs/Nb & QW & Sputtering & 630 & 52.60 & 2 & 0.10 \\
  \hline
  Takayanagi \textit{et al.} 1995 [\citenum{TAKAYANAGI1995}]  & InAs/Nb & QW & Sputtering & 400 & 20.80 & 3 & 0.05 \\
  \hline
  Giazotto \textit{et al.} 2004 [\citenum{giazotto_josephson_2004}] & InAs/Nb & QW & Sputtering & 190 & 11.00 & 15 & 0.12 \\
  \hline
  \multirow{4}{*}{Heida \textit{et al.} 1998 [\citenum{Heida98}]}
  & InAs/Nb & QW & Sputtering & 320 & 0.13 & 327 & 0.03 \\
  & InAs/Nb & QW & Sputtering & 470 & 0.24 & 324 & 0.06 \\
  & InAs/Nb & QW & Sputtering & 630 & 0.05 & 381 & 0.02 \\
  & InAs/Nb & QW & Sputtering & 780 & 0.10 & 431 & 0.03 \\
 \hline
  Zellekens \textit{et al.} 2020 [\citenum{Zellekens2020}] & InAs/Al & NW & Epitaxy & 75 & 0.02 & 3330 & 0.37 \\
  \hline
  Perla \textit{et al.} 2020 [\citenum{Perla2020}] & InAs/Nb & NW & Epitaxy & 55 & 0.08 & 2850 & 0.14 \\
  \hline
  Gharavi \textit{et al.} 2017 [\citenum{Gharavi_2017}] & InAs/Nb & NW & Sputtering & 170 & 0.10 & 3000 & 0.24 \\
 \hline
  \multirow{2}{*}{Mayer \textit{et al.} 2020 [\citenum{mayer_superconducting_2020}]} 
  & InAsSb/Al & QW & Epitaxy & 500 & 1.16 & 230 & 1.27\\
  & InAsSb/Al & QW & Epitaxy & 1000 & 0.57 & 491 &  1.33 \\
  \hline
  Ke \textit{et al.} 2019 [\citenum{ke_ballistic_2019}] & InSb/NbTiN & NW & Sputtering & 700 & 1.90 & 50 & 0.11 \\
  \hline 
  Wiedenmann \textit{et al.} 2015 [\citenum{wiedenmann_4-periodic_2016}] & HgTe/Nb & QW & Sputtering & 150 & 5.0 & 50 & 0.21\\
  \hline
  Pallecchi \textit{et al.} 2008 [\citenum{Pallecchi2008}] & CNT/NbPd & NW & Sputtering & 350 & 0.03 & 350 & 0.01 \\
  \hline
\end{tabular}
\caption{Comparison of our work on Josephson junction characterization and other reported characterization.}
\label{table:1}
\end{table*}

\section{Interface Transparency and Anharmonicity in JJ-FETs}

While a highly transparent interface is desired for the JJ-FET to be in the short ballistic regime, high transparency could have an adverse effect on qubit anharmonicity. This effect is detailed in the work of Kringh\o{}j \textit{et. al}~\cite{kringhoj_2018}. 

The anharmonicity is defined as the difference between the frequency corresponding to a transition from the ground state $\ket{0}$ to the first excited state $\ket{1}$ and the frequency corresponding to a transition from $\ket{1}$ to the second excited state $\ket{2}$. These frequencies are $\omega_0$ and $\omega_1$ respectively.

In the short junction limit~\cite{Beenaker1991} for a gatemon qubit in the transmon regime~\cite{Koch2007}, $\alpha$ takes the form, when expanded in powers of the phase difference across the junction $\hat \phi$ \cite{kringhoj_2018},

\begin{equation}
    \alpha = \omega_{0} - \omega_{1} = -E_C \left( 1 - \frac{3}{4} \frac{\sum_i T_i^2}{\sum_i T_i}\right),
\end{equation}

neglecting terms fourth order in $\hat \phi$ and larger. The transmission eigenvalues are $T_i$, where $i$ indexes the channel, and the charging energy is $E_C$. For a conventional SIS junction in which we take the limit where $T_i \rightarrow 0$, the anharmonicity becomes $\alpha = -E_C$. For a highly transparent interface with a single channel, the anharmonicity becomes $\alpha = -E_C/4$, with $E_C$ the charging energy. This is significantly reduced as compared to the case with little to no interface transparency, as is common with transmon superconductor-insulator-superconductor junctions~\cite{Koch2007}. 




We note that while these results show that the transparency must be optimized to maximize the qubit anharmonicity and supercurrent range of the JJ-FET, the optimum value of gate voltage or device geometry (few modes as in nanowires~\cite{kringhoj_2018} or many modes in wide junctions in our current case) requires tuning. The above analysis holds for a ballistic description of the JJ-FET near zero gate voltage while at the operating regime and these simplifying assumptions may not hold due to diffusive transport. Ideally the device will operate at single ballistic mode or will be many modes in a diffusive regime to achieve the inductance for the desired qubit frequency.


\section{One to two microwave photon transition}


There has been great progress in creating a gatemon style qubit by using a transmon qubit with its Josephson junction replaced by a JJ-FET~\cite{deLange2015}. Gatemon qubits first realized on InAs nanowires coated by a layer of epitaxial Al were successfully demonstrated~\cite{Larsen_PRL, CasparisBenchmarking}, showing coherence times $T_1$ up to 5 $\mu$s. However, nanowire based gatemon qubits present challenges for reproducibility when scaling the system, due to the necessity of a bottom up fabrication procedure. Many proposals have since focused on developing a gatemon qubit platform based on 2D semiconductor-superconductor heterostructures~\cite{Casparis2018}. The device can then be patterned utilizing widely available lithographic capabilities making this platform more amenable to scaling.


\begin{figure}[h]
    \centering
    \includegraphics[width=0.48\textwidth]{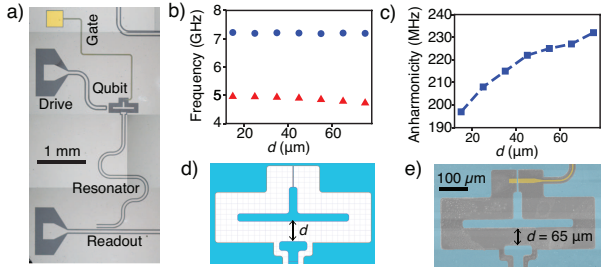}
    \caption{Images of Gatemon device and HFSS Simulations of coupling distance in the gatemon qubit-Al/InP system. (a) Stitched optical image of a gatemon style qubit showing the qubit, the qubit drive, electrostatic gate, the resonator, and a part of the readout transmission line for probing the resonator (b) Frequency of the qubit and the readout resonator as a function of the coupling distance $d$ between the qubit and the resonator. Red triangle and blue circle markers represent the qubit and readout resonator frequencies respectively. (c) Anharmonicity expected for the same range of coupling distances. (d) Geometry of the resonator and qubit designs with coupling distance $d$ indicated. (e) Corresponding false-colored SEM image of the resonator and qubit with a measured coupling distance $d = 65$ $\mu$m. }
    \label{fig:hfss}
\end{figure}

\begin{figure}[h!]
    \centering
    \includegraphics[width=0.48\textwidth]{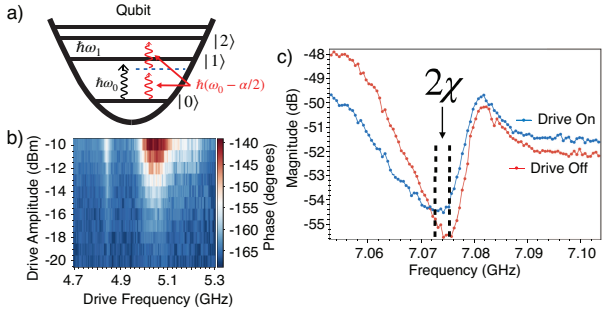}
    \caption{(a) Schematic of the energy levels in a superconducting qubit, detailing one and two photon excitations.  The single photon (left black) illustrates the transition from $\ket{0}$ to $\ket{1}$ with energy $\hbar\omega_0$.  The energy of transition from $\ket{1}$ to $\ket{2}$ with energy $\hbar\omega_1$ is also labeled.  The dashed line (blue) illustrates a virtual state utilized by the two right photons (right red) allowing for a two photon transition from $\ket{0}$ to $\ket{2}$. Each of the photons associated with this event have an energy of $\hbar (\omega_0 - \alpha/2)$. (b) Spectroscopy of phase of microwave transmission measurement on a gatemon sample shows one and two photon absorption at frequencies of $\sim 5.05$ GHz and $\sim 4.84$ GHz.  This implies an anharmonicity of $\alpha \sim 420$ MHz and coupling strength of $g \sim 153$ MHz. (c) Transmission for when the qubit drive is off and the qubit is in $\ket{0}$, (red) and on where the qubit is in $\ket{1}$ (blue), allowing us to extract a value for the dispersive shift $\chi$.}
    \label{fig:microwave}
\end{figure}

In this work, we design and fabricate a gatemon qubit to study the energy spectrum of our JJ-FET device. A stitched optical micrograph of the device can be seen in Fig.~\ref{fig:hfss}(a). The qubit consists of a JJ-FET with an effective capacitance in parallel. The qubit is then read through a capacitively coupled coplanar waveguide (CPW) resonator, which is subsequently coupled to a transmission line, along with a CPW resonator used for the drive line. We fabricate the gatemon qubit first by etching away the InAs-Al heterostructure in all regions except the active region of the junction noted by ``Qubit'' in Fig.~\ref{fig:hfss}(a).  We then blanket sputter a layer of Al over the whole device.  After this step we use E-beam lithography to pattern the microwave circuit into the sputtered Al, along with the junction gap into both the epitaxial and sputtered Al.  We then perform another round of lithographic patterning to define our gate contact and then follow a similar oxide, and gate deposition as mentioned previously.

We use Ansys, a High Frequency Structure Simulator (HFSS) software used for 3D electromagnetic modelling of the qubit and readout CPW resonator. The design of the CPW resonator and the coupling geometry were generated using PyEPR~\cite{minev2020energyparticipation} and is shown in Fig.~\ref{fig:hfss}(d). A corresponding false-colored scanning electron microscope image is seen in Fig.~\ref{fig:hfss}(e), with a coupling distance $d=65 \,\mu$m. We monitor how varying the coupling distance $d$ (and therefore the coupling strength $g$) affects the resonator's resonant frequency $\omega_r$, the qubit ground to first excited state transition frequency $\omega_0$ (hereafter called the qubit frequency), and the qubit anharmonicity, $\alpha$. These are shown in Fig.~\ref{fig:hfss}(b) and \ref{fig:hfss}(c). The resonant frequency $\omega_r$ of the readout cavity is $\sim$ 7 GHz, shown as blue circles. The qubit frequency $\omega_0$ is $\sim$ 5 GHz, shown as red triangles.  We vary the coupling distance $d$ between the qubit and resonator and find a range of qubit anharmonicities from $\alpha =$ 200-240 MHz for coupling distances of $d = $ 20 to 70 $\mu$m. One can see that qubit and resonator frequencies are largely unaffected by varying the coupling distance.


The qubit energy diagram is schematically shown in Fig.~\ref{fig:microwave}(a). The gatemon qubit has its potential energy landscape modelled by an anharmonic \textit{LC} oscillator~\cite{Koch2007, Schreier2008}. Detailed in Fig.~\ref{fig:microwave}(a) are the ground to first excite state transition, and the first to second excited state transition, characterized by their energy separations $\hbar \omega_0$ and $\hbar \omega_1$ respectively.  For an anharmonicity $\alpha = \omega_1-\omega_0$, a two photon transition from the ground state to the second excited state via an intermediate virtual state. Each photon participating in this transition have a frequency of $\omega_0-\alpha/2$. 

To analyze the anharmonicity, we probe the effect of the drive power on the readout spectrum. Figure~\ref{fig:microwave}(b) shows a sweep of the qubit drive power (drive amplitude) and frequency where color represents the magnitude of the transmission when probed at a fixed frequency of 7.08 GHz and on average about 50 photons.  We find the quality factors for the resonance shown in Fig. \ref{fig:microwave}(c) to be $Q_e \sim 850$ for the external quality factor and $Q_i \sim 850$ for the internal quality factor, found through a circle-fitting algorithm applied to the complex transmission signal~\cite{Probst}.  We note that the low quality factors of the readout resonator can be attributed to limitations from dielectric loss in the InP substrate~\cite{Casparis2018}, difficulty in cleaning the InP oxide, and fabrication non-idealities. When the qubit is driven with high power from $\ket{0}$ to $\ket{1}$, the transition is quite broadened. Similarly we observe a $\ket{0}$ to $\ket{2}$ transition as a result of a two-photon process~\cite{kringhoj_2018} where the frequency of each photon is $(\omega_0+\omega_1)/2$. Figure ~\ref{fig:microwave}b shows these absorptions corresponding to frequencies of $5.05$ GHz and $4.84$ GHz.  The single photon excitation process is simply the qubit transitioning from $\ket{0}$ to $\ket{1}$.  The two photon excitation is one which goes directly from $\ket{0}$ to $\ket{2}$.  This procedure is one where two photons are absorbed from the drive simultaneously at a frequency which when summed provide the correct amount of energy for this jump.  Given the nature of the experiment we are providing a single tone on the drive line and notice as expected that this condition is driven probabilistically at coincidental moments where two photons arrive close enough to be absorbed simultaneously. As power decreases the signal weakens rapidly as there are less and less photons available. The transition takes place by use of a virtual state as diagrammed in Fig.~\ref{fig:microwave}(a) as a dashed line slightly below $\ket{1}$ with two photons shown driving the transition. This implies an anharmonicity of $\alpha = 420 \pm 51$ MHz. It is known that there is a dispersive shift $2\chi$ in the readout resonator frequency whether the qubit is in the $\ket{0}$ or $\ket{1}$ state. Figure~\ref{fig:microwave}(c) shows the readout measurement when the qubit drive amplitude is on and off. For this device we find $\chi \sim 1.5$ MHz. Knowing $\chi$, $\alpha$ and $\delta = \abs{\omega_r - \omega_0} = 2.024 \pm .025$ GHz we can estimate the coupling strength of $g = \sqrt{\chi \delta (\alpha + \delta)/\alpha} = 130.7 \pm 32.9$ MHz. These values are comparable with typical transmon circuits \cite{Koch2007} which suggest that JJ-FETs are a promising candidate for gatemon qubits.

Unlike in nanowire gatemon qubits the fixed frequency transmon qubits. This suggests that the control can be as fast for Gatemon qubits. In addition, our coupling strength is similar to that of the gatemon fabricated on a 2D platform~\cite{Casparis2018} which suggests that the overall nonlinearlity is increased in JJ-FETs based on 2DEGs compared to those on nanowires. However there are materials considerations including the choice and thickness of the top barrier. In our case a 10~nm In$_{0.81}$Ga$_{0.19}$As was chosen to increase the mobility and the strong coupling to the Al layer. From our studies, it seems a thicker or a higher band gap like In$_{0.81}$Al$_{0.19}$As would be a better choice in reducing the transparency and the associated supercurrent in the junction. 



\section{Conclusion}


A 2D platform for gate voltage tunable JJ-FET circuit elements offers exciting possibilities for gatemon qubits and other tunable microwave circuits. We leverage advancements in the growth of high mobility InAs quantum wells contacted with Al and design to yield JJ-FETs that demonstrate a wide range of tunability using simply a gate voltage. We also report on the measurement of anharmonicity and coupling strength in a gatemon style qubit fabricated on this InAs-Al heterostructure, noting how these may be affected by design and materials parameters, such as the barrier composition and thickness, junction length, doping density, and coupling distance.  

As an SiO$_x$ or AlO$_x$ gate dielectric could introduce a higher density of charge traps and interface states~\cite{chobpattana_nitrogen-passivated_2013}, it may be more favorable to use monolayer thick h-BN as a gate dielectric, which has low density of charge traps, and low microwave absorption, possibly leading to better device performance while maintaining supercurrent tunability~\cite{barati_tuning_2021}.


\section{Acknowledgements}
We acknowledge support from US Army Research Office under agreement W911NF1810067. Joseph Yuan acknowledges funding from the ARO/LPS QuaCGR fellowship reference W911NF1810067. This work was performed in part at the Nanofabrication Facility at the Advanced Science Research Center at The Graduate Center of the City University of New York.

\bibliography{References_Shabani_Growth}
\pagebreak

\end{document}